\documentclass[aps,prl,showpacs,preprint,superscriptaddress]{revtex4}
\usepackage{epsfig}
\begin{document}

\title{Extreme nuclear shapes examined via Giant Dipole Resonance lineshapes in hot light mass system}

\author{Deepak Pandit}
\affiliation{Variable Energy Cyclotron Centre, 1/AF-Bidhannagar, Kolkata-700064, India}

\author{S. Mukhopadhyay}
\affiliation{Variable Energy Cyclotron Centre, 1/AF-Bidhannagar, Kolkata-700064, India}

\author{Srijit Bhattacharya}
\affiliation{Department of Physics, Darjeeling Government College, Darjeeling-734101, India}

\author{Surajit Pal}
\affiliation{Variable Energy Cyclotron Centre, 1/AF-Bidhannagar, Kolkata-700064, India}

\author{A. De}
\affiliation{Department of Physics, Raniganj Girls' College, Raniganj-713358, India}

\author{S. Bhattacharya}
\affiliation{Variable Energy Cyclotron Centre, 1/AF-Bidhannagar, Kolkata-700064, India}

\author{C. Bhattacharya}
\affiliation{Variable Energy Cyclotron Centre, 1/AF-Bidhannagar, Kolkata-700064, India}

\author{K. Banerjee}
\affiliation{Variable Energy Cyclotron Centre, 1/AF-Bidhannagar, Kolkata-700064, India}

\author{S. Kundu}
\affiliation{Variable Energy Cyclotron Centre, 1/AF-Bidhannagar, Kolkata-700064, India}

\author{T. K. Rana}
\affiliation{Variable Energy Cyclotron Centre, 1/AF-Bidhannagar, Kolkata-700064, India}

\author{A. Dey}
\affiliation{Variable Energy Cyclotron Centre, 1/AF-Bidhannagar, Kolkata-700064, India}

\author{G. Mukherjee}
\affiliation{Variable Energy Cyclotron Centre, 1/AF-Bidhannagar, Kolkata-700064, India}

\author{T. Ghosh}
\affiliation{Variable Energy Cyclotron Centre, 1/AF-Bidhannagar, Kolkata-700064, India}

\author{D. Gupta}
\affiliation{Variable Energy Cyclotron Centre, 1/AF-Bidhannagar, Kolkata-700064, India}
\affiliation{Bose Institute, Department of Physics \& Centre for Astrophysics \& Space Science, Salt Lake City, Kolkata - 700091, India}

\author{S. R. Banerjee}
\email[e-mail:]{srb@veccal.ernet.in}
\affiliation{Variable Energy Cyclotron Centre, 1/AF-Bidhannagar, Kolkata-700064, India}


\date{\today}

\begin{abstract}
The influence of alpha clustering on nuclear reaction dynamics is investigated using the giant dipole resonance (GDR) lineshape studies in the reactions $^{20}$Ne ($E_{\text{lab}}$=145,160 MeV) + $^{12}$C  and $^{20}$Ne ($E_{\text{lab}}$=160 MeV) + $^{27}$Al , populating $^{32}$S and $^{47}$V, respectively. The GDR lineshapes from the two systems are remarkably different from each other. Whereas, the non alpha-like $^{47}$V undergoes Jacobi shape transition and matches exceptionally well with the theoretical GDR lineshape estimated under the framework rotating liquid drop model (RLDM) and thermal shape fluctuation model (TSFM) signifying shape equilibration, for alpha cluster $^{32}$S an extended prolate kind of shape is observed. This unusual deformation, seen directly via $\gamma$-decay for the first time, is predicted to be due to the formation of orbiting di-nuclear configuration or molecular structure of $^{16}$O+ $^{16}$O in $^{32}$S superdeformed  band.

\end{abstract}
\maketitle

In recent years, there have been intense theoretical and experimental efforts \cite{intro1,beck1}, to search for highly (super/hyper) deformed (SD/HD) nuclear systems. There are indications that such 
highly deformed shapes are likely to be observed in light $\alpha$-like systems 
($A_{CN}$ $\sim$ 20-60) at higher angular momenta (typically, $\gtrsim$ 15$\hbar$) and 
excitation energies (typically, $\gtrsim$ 40 MeV). Experimentally, large deformations
were observed in $^{36}$Ar \cite{intro2}and $^{40}$Ca \cite{intro3}, where the 
deformations were studied using $\gamma$-spectroscopic techniques. In recent 
times, experimental indications of hyperdeformation have also been 
reported in the decay of $^{56}$Ni \cite{intro5,von1} and $^{60}$Zn \cite{intro6}, 
where charged particle spectroscopy was used to identify the ternary fission-like 
decay of hyperdeformed composites. Hyperdeformed shape in $^{36}$Ar has also 
been predicted \cite{intro7}, however, it is not yet firmly 
established exprimentally \cite{intro1}. 

Such large deformations observed in light $\alpha$-like systems are 
believed to be due to the  occurence of either quasimolecular resonances or nuclear 
orbiting \cite{intro4}, which have the origin in $\alpha$-cluster structure of 
these nuclei. On the other hand, rapidly rotating light nuclei in general are 
likely to undergo Jacobi shape transition at an angular momentum value near the fission 
limit, where the shape changes abruptly from non-collective oblate to collective 
triaxial / prolate shape. Since it is possible for the light mass nuclei to attain very high
angular velocities beyond the critical point for Jacobi transition
without undergoing fission, existence of exotic shapes in nuclei with a large deformation becomes likely.
Signatures of such shape transitions in $^{45}$Sc\cite{ksin} 
and $^{46}$Ti\cite{intro8,brek,kmie}  have been reported from the study of line shapes of giant dipole resonance (GDR) built on excited states. Recently, the Jacobi shape transition has been confirmed for
$^{48}$Cr\cite{sand} from high-spin spectroscopy and corresponding theoretical calculation 
using the LSD liquid drop model \cite{dudek}. It is, therefore, relevant 
to explore the relationship between the shapes of the light $\alpha$-like systems 
and the corresponding Jacobi shapes, which would help in understanding the reaction 
dynamics of light $\alpha$-like systems and the role played by $\alpha$-clustering 
in determining the shape.    

The aim of the present study is to make quantitative experimental estimation of the 
deformed shapes of light alpha and non-alpha  systems using GDR lineshape 
studies and compare them with the corresponding predictions for equilibrium Jacobi 
shapes. For $\alpha$-like system, we have taken $^{20}$Ne + $^{12}$C system and for 
non $\alpha$-like system, we have chosen $^{20}$Ne + $^{27}$Al system. 
The $^{20}$Ne + $^{12}$C system  is well established orbiting system 
\cite{intro9}, and our previous studies on enhancement of fragment yield near the 
entrance channel \cite{intro10} as well as $\alpha$-spectroscopic studies  
\cite{intro11} have strongly indicated a highly deformed orbiting dinuclear shape 
for this system. On the contrary, for  non-alpha like $^{47}$V system, it is well
established that there is no significant entrance channel effect and fusion-fission 
compound nuclear yield is dominant in accordance with the qualitaitve expectation
from the number of open channels model \cite{ray,beck2}. 
Since $\alpha$-spectroscopic studies can only indicate effective deformation in
an indirect way, it is worthwhile to complement the above studies with the investigation of nuclear 
shapes of the excited rotating systems through the GDR $\gamma$-decays in a more direct manner.

The GDR, linear oscillations of the protons and neutrons in the hot nucleus, 
occurs on a time scale that is sufficiently short. Thus, these can compete with other 
modes of nuclear decay. In addition, the resonance couples directly with the other
nuclear degrees of freedom, such as shape 
degrees of freedom, and thus can provide information on the shape evolution of the nuclei at finite 
temperature and fast rotation. The resonance energy being proportional to the inverse 
of the nuclear radius, the GDR strength function splits in the case of deformed nuclei 
and the investigation of the strength distribution gives a direct access to nuclear deformations. 
Here, we report the experimental GDR strength functions for the hot, rotating composites formed in 
$^{20}$Ne + $^{12}$C and $^{20}$Ne + $^{27}$Al systems and show, {\it for the first time}, 
that, whereas the experimentally extracted shape for $^{20}$Ne + $^{27}$Al conforms 
to the predicted jacobi shape, extracted shape of the $^{20}$Ne + $^{12}$C system 
is highly elongated (prolate) and does not conform to the corresponding Jacobi shape, 
clearly highlighting the difference between the reaction mechanisms in the two cases.

\begin{figure}
\begin{center}
\includegraphics[height=4.0 cm, width=8.0 cm]{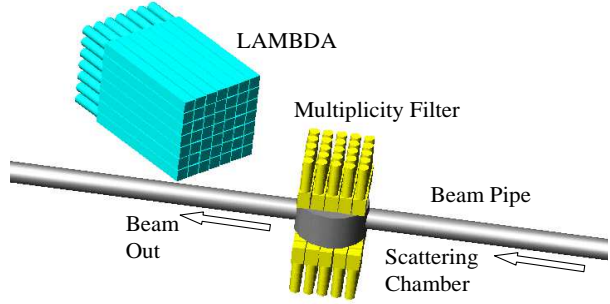}
\caption{\label{setup} (color online) Schematic view of experimental setup for the LAMBDA
(Large BaF2 Array) spectrometer in a 7 $\times$ 7 matrix arrangement along
with the low energy $\gamma$-ray multiplicity filter.}
\end{center}
\end{figure}

\begin{figure}
\begin{center}
\includegraphics[height=5.0 cm, width=5.0 cm]{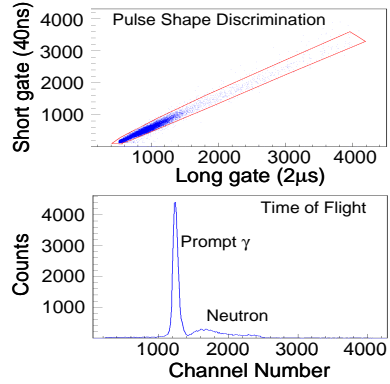}
\caption{\label{psd}(color online) The experimental pulse shape discrimination 
and time of flight spectra obtained from a single detector.}
\end{center}
\end{figure}

The $^{47}$V and $^{32}$S nuclei were formed by bombarding pure 
1 mg/cm$^2$ thick $^{27}$Al and $^{12}$C targets with accelerated
$^{20}$Ne beams from the K-130 cyclotron at the Variable Energy 
Cyclotron Centre, Kolkata, India. The initial excitation energy
for $^{47}$V compound system was $E_x$=108 MeV corresponding to a 
projectile energy $E_{\text{proj}}$=160 MeV. Similarly the initial 
excitation energies for $^{32}$S nucleus were $E_x$=73 \& 78 MeV 
corresponding to the projectile energies of $E_{\text{proj}}$=145 \& 160 MeV.
The corresponding critical angular momenta for $^{47}$V and $^{32}$S
nuclei are $l_{cr}$=38$\hbar$ and 24$\hbar$ respectively, same as those
extracted from previous complete fusion mesurement by using the sharp 
cut-off approximation for $^{47}$V \cite{vansen} and $^{32}$S \cite{saint}.
These values extend well beyond the critical angular momentum values
of 29.6$\hbar$ and 21.5$\hbar$ at which the Jacobi transitions are
predicted to occur for these nuclei(according to systematics $J_c=1.2A^{5/6}$ \cite{kus1}). 
They are formed in identical
conditions as in the previous charged particle experiment reported
earlier \cite{intro11}. The experimental arrangement and technique
was similar to that described earlier \cite{Srij}. The high energy
photons were detected at $\theta_\gamma$=55$^\circ$ with a part
of the LAMBDA (Large Area Modular BaF$_2$ Detector Array) spectrometer
arranged in a 7 $\times$ 7 square matrix \cite{Supm}. 
An event-by-event information of the populated
angular momentum was recorded using a 50 element BaF$_2$ based
low energy $\gamma$-multiplicity filter in coincidence with the high energy photon events. 
The multiplicity filter was kept at a distance of 10 cm from the target which
ensured the selection of higher part of the spin distribution.
The schematic view of the experimental setup is shown in Fig.\ref{setup}.
Time of flight(TOF) technique was used to eliminate neutrons while pulse shape discrimination (PSD)
was adopted to reject pile-up events for the individual detector elements.
The PSD and the TOF spectra for a single detector is shown in Fig.\ref{psd}
clearly indicating excellent rejection of neutrons and pile-up events.

\begin{figure}
\begin{center}
\includegraphics[height=9.5 cm, width=8.5 cm]{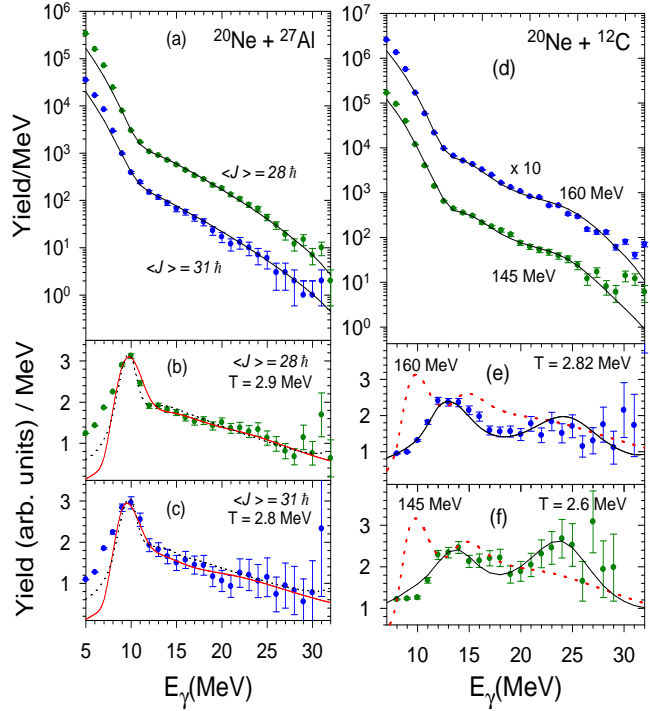}
\caption{\label{spec} (color online) (a)The experimental $\gamma$-spectra
and extracted 
linearized GDR strength functions (b,c) for $^{47}$V for the 
two angular momentum windows. The 
dotted lines are the respective CASCADE predictions and the solid lines are 
the predicted line shapes from the free energy minimazation technique.
Same for $^{32}$S for two incident energies (d,e,f). The solid 
lines are the CASCADE predictions using 2-GDR components and the dotted 
lines are the predicted line shape from the free energy calculations.}
\end{center}
\end{figure}

The linearized GDR spectra from $^{47}$V (for two ang. mom. windows) and $^{32}$S 
(for two incident energies) are shown in Fig.\ref{spec}.
The GDR lineshapes from the two systems are remarkably different, from, 
which one usually gets in the case of a spherical or a near
spherical system and indicate large deformations.
The most striking feature in the case of $^{47}$V populated at 
28$\hbar$ \& 31$\hbar$ is the strong enhancemant in $\gamma$-yield at 
$\sim$10 MeV similar to one observed in $^{46}$Ti \cite{intro8}
earlier. It is characteristic of a large deformation and 
the effect of coriolis splitting
due to very high angular velocity in the system. Such very high
angular velocities are usually achieved by the system normally
beyond the Jacobi transition point.
However, for $^{32}$S no such enhancement is seen though the
nucleus is populated at spins well beyond Jacobi transition point. 
A 2-component GDR strength 
function fits the experimental data fairly well. 
The extracted parameters are given in table-\ref{tab:para2}.
The shape looks more like a 
highly extended prolate and is seen for the first time for this nucleus.
The GDR lineshape for $^{47}$V, on the other hand, is more 
complex and a simple two or three component GDR strength function 
fails to describe the experimental observations. 
The linearized GDR lineshapes were extracted using a modified version
of the statistical model code CASCADE \cite{Srij,pul}. For both the cases we have adopted the 
Ignatyuk level density prescription \cite{igna} keeping the 
asymptotic level density parameter $\widetilde{a}$ = A/8 MeV$^{-1}$. 
The CASCADE calculations have been performed with the same parameters as
used in the charged particle analysis except for the $\delta1$ parameter in the case $^{32}$S. 
We have taken $\delta1$ as 1.0 $\times$ 10$^{-3}$ in order to explain the 
low energy part of the $\gamma$ spectrum.

In order to interpret the extracted GDR strength functions in the
entire $\gamma$-ray energy region (5-32 MeV) and to understand the
equilibrium deformations in these hot and rotating nuclear systems,
a calculation is performed for estimating the equilibrium shape of
a nucleus by minimizing the total free energy under the framework of 
rotating liquid drop model (RLDM) and thermal shape fluctuation model (TSFM)
for a given temperature $T$ and angular momentum $J$. Earlier, a similar
model with a modified liquid drop parametrization (LSD Model) was used to extract the shape 
evolution of $^{46}$Ti and $^{48}$Cr \cite{intro8,sand,dudek}.

\begin{figure}
\begin{center}
\includegraphics[height=7.0 cm, width=8.3 cm]{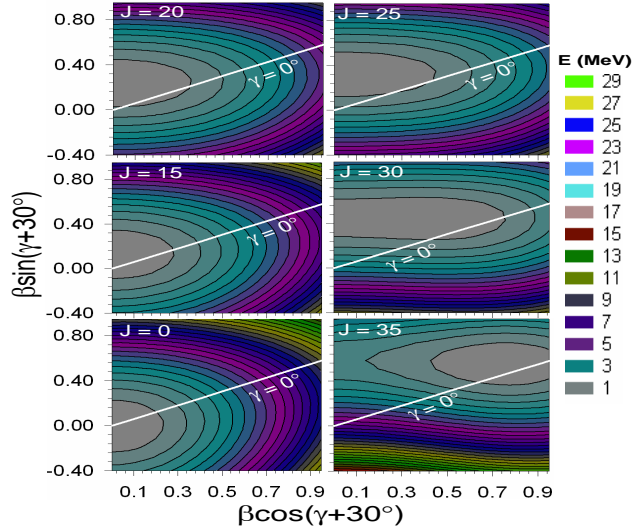}
\caption{\label{cont}(color online) Liquid Drop free energy surfaces at different spins for $^{47}$V . The line represents the prolate shape ($\gamma$=0).}
\end{center}
\end{figure}

\begin{figure}
\begin{center}
\includegraphics[height=4.0 cm, width=8.5 cm]{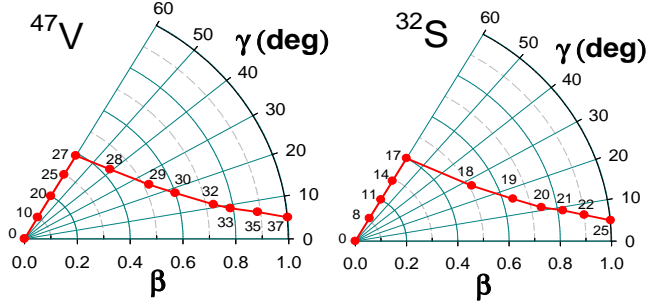}
\caption{\label{bg}(color online) The equilibrium shapes are plotted as a function 
of quadrupole deformation parameters $\beta$ and $\gamma$ for
different spins for $^{47}$V and $^{32}$S nuclei. 
The discrete spin values are represented alongside 
the data points.}
\end{center}
\end{figure}

The free energy for a hot rotating nucleus at a constant spin ($J$),
in a liquid drop picture, can be written as \cite{Alhad2}, 
\begin{equation}
	F(T,J,\beta,\gamma)= E_{LDM}(\beta,\gamma) -TS + 		
												\frac{J(J+1)\hbar^2}{2(\omega.I.\omega)}
\end{equation}
where
\begin{eqnarray}
\omega.I.\omega = I_{xx}sin^2\theta cos^2\phi+ I_{yy}sin^2\theta sin^2\phi + I_{zz}cos^2\theta \nonumber
\end{eqnarray}
is the moment of inertia about the rotation axis $\omega$.
I$_{xx}$, I$_{yy}$, I$_{zz}$ are the principal rigid body moments of inertia
and S is the entropy of the system.
The dependence of level densities on deformation 
and shell corrections are assumed to be small and neglected
as the nuclear temperatures in this case are $\sim$3 MeV.
E$_{LDM}$($\beta$,$\gamma$) is the deformed liquid 
drop energy, calculated in terms of $\beta$ and $\gamma$, 
the intrinsic quadrupole deformation parameters 
\cite{Will, Moll, Dave, Krap}. 
The deformed moment of
inertia can be written in Hill-Wheeler parametrization as,
$I_{zz}=B_1 \frac{2}{5} m r_0^2 A^{5/3}$, where,
$B_1=\frac{1}{2} ({r_1}^2+{r_2}^2)$, 
$r_1 = R_x/R_0 = \exp{\left(\sqrt{5/4\pi}\right)} 
\beta \cos{\left(\gamma-2\pi/3\right)}$
and, $r_2 = R_y/R_0 = \exp{\left(\sqrt{5/4\pi}\right)} 
\beta \cos{\left(\gamma+2\pi/3\right)}$. $R_x$, $R_y$ are
the deformed radius parameters along the directions perpendicular
to the spin axis and $R_0$ is that for an equivalent undeformed spherical
shape. We have considered rotation axis along z, different from the common convention of $\omega$ as parallel
to $x$, as it leads to simplification of many expressions and calculations. However, we have plotted
 $\gamma$ as ($\gamma$ - 120$^{\circ}$) in Fig. \ref{cont} and Fig. \ref{bg} in order to represent
the equilibrium deformation in the conventional way.

\begin{table}[htbp]
	\centering
	\caption{Extracted GDR parameters for $^{32}$S used for CASCADE 
	calculations. $E_1, E_2, \Gamma_1, \Gamma_2, S_1, S_2$ are the
	resonance energies, widths and fractional strengths respectively
	for the two GDR components.}
	\label{tab:para2}
   \begin{ruledtabular}
		\begin{tabular}{cccccccc}
			$E_{\text{proj}}$ & $E_1 $ & $\Gamma_1 $ & $S_1$ & 
			$E_2 $ & $\Gamma_2 $ & $S_2$ & $\beta$ \\
			(MeV) & (MeV) & (MeV) & & (MeV) & (MeV) & &      \\ \hline
			 145  &  14.5 &  6.2  & 0.37 & 25.4 & 7.5 & 0.63 & 0.68 \\
			 160  &  14.0 &  6.2  & 0.32 & 26.0 & 8.2 & 0.68 & 0.76 \\
		\end{tabular}
   \end{ruledtabular}
\end{table}

The free energy surfaces were obtained for different spins over
the entire $\beta$-$\gamma$ space as shown in Fig.\ref{cont} for 
$^{47}$V compound nucleus. It can be clearly seen that as the spin
increases, the minima of the free energy surfaces move towards
increasing oblate deformation (along y-axis) and 
suddenly make a transition towards the prolate ($\gamma$=0$^\circ$) 
deformation (represented by the line) at high spins. The gradual
evolution in shape and the transition point are evident 
in a polar $\beta$-vs-$\gamma$ plot for discrete spins (Fig.\ref{bg}). 
It is clear from the figure that for 
$^{47}$V, with the increase in angular velocity, the nucleus becomes more 
and more oblate deformed ($\gamma$=60$^\circ$). 
After a critical spin of 27$\hbar$ it suddenly becomes triaxial
(60$^\circ$ $< \gamma <$ 0$^\circ$) and approaches a prolate shape
at still higher spins.

Experimentally, the average angular momentum was deduced from a
selection of the fold distribution of the low energy $\gamma$-
multiplicity filter array \cite{Srij} and corresponding to the 
$<J>$ values the equilibrium shape of the nucleus is known 
(Fig.\ref{bg}). The GDR observables ($E_{GDR}$, $\Gamma_{GDR}$) 
are calculated built on these shapes using Hill-Wheeler parametrization 
and considering $\Gamma^i_{GDR}=\Gamma_0 (E^i_{GDR}/E_0)^\delta$ \cite{carlos},
where, $\Gamma_0$ is the ground state GDR width and $\delta$ is
taken as 1.9. This procedure in general gives three different
GDR frequencies (along the three unequal axes) and 
widths corresponding to a particular 
angular momentum and equilibrium shape of the nucleus.
These three GDR frequencies further split (the ones perpendicular
to the spin axis) due to coriolis effect as the GDR vibrations
in a nucleus couple with its rotation when viewed from a non-
rotating frame \cite{Neer}. The magnitude of the split depends 
on the magnitude of the rotation frequency and give rise
to five GDR components. The resultant GDR lineshape is obtained 
as a superposition all the components ($E_i$, $\Gamma_i$),
\begin{eqnarray}
\sigma_{total}=\sum^5_{i=1} \frac{E^2_\gamma \Gamma_i}
{(E^2_\gamma-E_i^2)^2 + E^2_\gamma \Gamma_i^2} \nonumber
\end{eqnarray}

The TSFM assumes that at high T and J, the GDR vibration samples an ensemble
of different shapes around the equilibrium shape of the compound nucleus.
Thus, the GDR lineshapes are generated (according to
the adiabatic TSFM \cite{Alhad1})
by averaging the GDR vibrations over the free energy
surfaces (using a Boltzman probability distribution
$\exp(-F/T)$) in the entire deformation space 
including full orientation fluctuation using,
\begin{eqnarray}
	\langle \sigma \rangle = \frac{\int D_\alpha e^{-F/T}
	(\omega.I.\omega)^{-3/2} \sigma}
	{\int D_\alpha e^{-F/T}(\omega.I.\omega)^{-3/2}} \nonumber
\end{eqnarray}
where, $D_\alpha=\beta^4 \sin(3\gamma) d\beta d\gamma d\Omega$ is the 
elemental volume.

The resultant lineshape is compared with the experimental data
and is plotted in Fig.\ref{spec} (solid line). It describes
the data for $^{47}$V remarkably well for both the experimetally measured 
spin windows of 28$\hbar$ \& 31$\hbar$ (spin uncertainties of $\pm$7$\hbar$ \& $\pm$8$\hbar$ respectively) 
at corresponding temperatures of 2.9 \& 2.8 MeV respectively. 
The presence of the enhancement in the lineshape at $\sim$10 MeV 
and the goodness of description are characteristic signatures of 
Jacobi transition in the case of $^{47}$V nucleus at these spin 
values.

\begin{table}[htbp]
	\centering
	\caption{Calculated GDR parameters for $^{47}$V from RLDM \& TSFM 
	calculations. $E_i, \Gamma_i, S_i$ are the
	resonance energies, widths and fractional strengths respectively
	for the five GDR components after coriolis splitting.}
	\label{tab:para1}
   \begin{ruledtabular}
		\begin{tabular}{c|cccc|cccc}
			$E_{\text{proj}}$ & 
			\multicolumn{4}{c|}{$<J>$=28$\hbar$} & 
			\multicolumn{4}{c}{$<J>$=31$\hbar$} \\
			(MeV) & $E_i$ &  $\Gamma_i$ & $S_i$ & $\beta$ &
			        $E_i$ &  $\Gamma_i$ & $S_i$ & $\beta$ \\ \hline
			      &  9.9  &  3.0  & 0.33 &      & 9.9  & 3.0  & 0.33 &  \\
			      &  14.5 &  5.3  & 0.15 &      & 14.1 & 5.1  & 0.15 &  \\
			 160  &  18.3 &  8.1  & 0.17 & 0.43 & 18.4 & 8.4  & 0.17 & 0.67 \\
			      &  23.1 &  11.3 & 0.17 &      & 23.0 & 11.5 & 0.17 &  \\
			      &  27.3 &  15.5 & 0.18 &      & 27.8 & 15.8 & 0.18 &  \\
		\end{tabular}
   \end{ruledtabular}
\end{table}

Since it is difficult to extract the angular momentum for very low mass accurately,
the $^{32}$S nucleus was populated with two incident energies, 145 and 160 MeV,
in order to populate $^{32}$S at different spins. The high fold gated spectrum of
$^{32}$S for both the energies is shown in Fig.\ref{spec}. 
The similar free energy calculations are done at both low and high spins, but it fails miserably 
to describe the experimental data. The calculation performed at 22$\hbar$ is shown in Fig.\ref{spec}
(dotted line (e),(f)).
Though the nucleus is populated well above the critical spin for Jacobi transition 
(Fig.\ref{bg}), it does not show the characteristic behaviour of such a 
transition in shape. Instead, the data is reproduced quite well
by a statistical model (CASCADE) calculation with a 2-component
GDR strength function (table-\ref{tab:para2}). 
The shapes suggest a strongly prolate
deformed nucleus ($\beta \approx$ 0.76 for $E_{\text{proj}}$=160 MeV, 
corresponding to an axis ratio of 1.7:1).
The $^{20}$Ne + $^{12}$C system was studied earlier at 5 MeV/u and 9.5 MeV/u 
to study the isospin effects in light mass nuclei \cite{kic} 
but only preliminary result was presented for the higher incident energy 
indicating towards incomplete fusion. Indeed, the pre-equilibrium component 
at 200 MeV incident energy, as measured from our earlier charge particle experiment, 
is around 50 $\%$ . However, the pre-equilibrium component in our case,
less than 10$\%$ for 160 MeV  
and negligible at 145 MeV, does not seem to have a major influence on the high energy $\gamma$-spectra.

The occurrence of such a large deformation without showing 
the characteristics of Jacobi transition is possible only 
if some other reaction mechanism is responsible. 
One of the possibilities could be the formation of 
an orbiting di-nuclear complex where the nucleus is 
not fully equilibrated (in terms of shape degrees of 
freedom) and maintains the entrance channel shape before 
finally splitting into two parts. Our charged particle 
studies have also indicated a highly deformed orbiting dinuclear
shape of this system. However, it can also be 
conjectured that the observed unusual deformation 
can be due to the formation of molecular structure 
of $^{16}$O + $^{16}$O cluster  in $^{32}$S. 
In the theoretical work of Kimura and Horiuchi \cite{kim},
it was predicted that the SD states of $^{32}$S have considerable 
amount of  $^{16}$O + $^{16}$O components and become more 
prominent as the excitation energy increases. 
The extracted deformation for two touching $^{16}$O was found to 
be $\beta$=0.73 which is in agreement with the experimentally 
extracted deformation from the resonance energy peaks. 
The occurrence of GDR in nuclei, where the entire nucleus takes part
in a collective manner, is clearly an effect of the mean field structure
of the nucleus. It is also known that other inherent structures 
(molecular resonance and/or orbiting di-nuclear complex) in light
$\alpha$-cluster nuclei may coexist with the mean field description of the nucleus.
Whether the experimental signatures of the overall nuclear deformation via GDR
$\gamma$-decay is due to the coexistence of these effects needs to be investigated 
further and are beyond the scope of this present study.

In summary, we have populated the nuclei $^{47}$V and $^{32}$S
at the highest spins and high excitations and studied their shapes
directly by looking at the lineshapes of their GDR decay. Both the
nuclei show highly deformed structures corresponding to highly fragmented GDR strength functions. 
The $^{47}$V
nucleus shows Jacobi triaxial shapes well beyond the critical
transition point. The evolution of the nucleus as a function of
angular momenta is estimated in the RLDM \& TSFM framework and found
to match exactly with the experiment. In the case of alpha cluster $^{32}$S nucleus, 
a highly deformed extended prolate configuration
is evident which does not follow the usual evolution of shape with angular momentum.
This unusual deformation, seen directly for the first time, can be speculated due 
to the formation of either orbiting di-nuclear configuration or
molecular structure of $^{16}$O + $^{16}$O in $^{32}$S SD band.

\end{document}